\documentclass[aps,prb,showpacs,superscriptaddress,twocolumn]{revtex4-1}

\usepackage{amsmath,graphics}
\usepackage[next]{inputenc}
\usepackage[dvips]{epsfig}
\usepackage[colorlinks=true,citecolor=blue,linkcolor=blue]{hyperref}
\usepackage{bbm}
\usepackage{bbold}
\usepackage{booktabs}
\usepackage{multirow}
\usepackage{hhline}
\usepackage{graphicx}
\usepackage{amsmath}
\usepackage{multirow}
\usepackage[usenames, dvipsnames]{color}
\usepackage{color,soul}
\usepackage{latexsym}

\usepackage{subfigure}
\usepackage{hyperref}
\usepackage{float}
\usepackage{amssymb}

\begin{document}

\title{Multiple Dirac Nodes and Symmetry Protected Dirac Nodal Line in Orthorhombic $\alpha$-RhSi}

\author{Shirin Mozaffari}\thanks{These authors contributed equally to this work}
\affiliation{National High Magnetic Field Laboratory, Florida State University, Tallahassee, Florida 32310, USA}

\author{Niraj Aryal}\thanks{These authors contributed equally to this work}
\affiliation{National High Magnetic Field Laboratory, Florida State University, Tallahassee, Florida 32310, USA}
\affiliation{Department of Physics, Florida State University, Tallahassee, Florida 32306, USA}

\author{Rico Sch\"{o}nemann}
\affiliation{National High Magnetic Field Laboratory, Florida State University, Tallahassee, Florida 32310, USA}

\author{Kuan-Wen Chen}
\affiliation{National High Magnetic Field Laboratory, Florida State University, Tallahassee, Florida 32310, USA}
\affiliation{Department of Physics, Florida State University, Tallahassee, Florida 32306, USA}

\author{Wenkai Zheng}
\affiliation{National High Magnetic Field Laboratory, Florida State University, Tallahassee, Florida 32310, USA}
\affiliation{Department of Physics, Florida State University, Tallahassee, Florida 32306, USA}

\author{Gregory T. McCandless}
\affiliation{Department of Chemistry and Biochemistry, The University of Texas at Dallas, Richardson, Texas 75080, USA}

\author{Julia Y. Chan}
\affiliation{Department of Chemistry and Biochemistry, The University of Texas at Dallas, Richardson, Texas 75080, USA}

\author{Efstratios Manousakis}\email{stratos@physics.fsu.edu}
\affiliation{National High Magnetic Field Laboratory, Florida State University, Tallahassee, Florida 32310, USA}
\affiliation{Department of Physics, Florida State University, Tallahassee, Florida 32306, USA}
\affiliation{Department   of    Physics,   University    of   Athens, Panepistimioupolis, Zografos, 157 84 Athens, Greece}

\author{Luis Balicas}\email{balicas@magnet.fsu.edu}
\affiliation{National High Magnetic Field Laboratory, Florida State University, Tallahassee, Florida 32310, USA}
\affiliation{Department of Physics, Florida State University, Tallahassee, Florida 32306, USA}

\begin{abstract}
  Owing to their chiral cubic structure, exotic multifold topological excitations have been predicted and recently observed in transition metal silicides like $\beta$-RhSi.
  Herein, we report that the topological character of RhSi is also observed in its orthorhombic $\alpha$-phase which displays multiple types of Dirac nodes very close to the Fermi level ($\varepsilon_F$) with the near absence of topologically trivial carriers. We discuss the symmetry analysis, band connectivity along high-symmetry lines using group representations, the band structure, and the nature of the Dirac points
  and nodal lines occurring near $\varepsilon_F$. The de Haas-van Alphen effect (dHvA) indicates a Fermi surface in agreement with the calculations.
  We find an elliptically-shaped nodal line very close to $\varepsilon_F$ around and near the $S$-point on the $k_y-k_z$ plane that results from the intersection of
  two upside-down Dirac cones. The two Dirac points of the participating Kramers degenerate bands are only 5 meV apart, hence an accessible magnetic field might induce
  a crossing between the spin-up partner of the upper-Dirac cone and the spin-down partner of the lower Dirac cone, possibly explaining the anomalies observed in the magnetic torque.
\end{abstract}

\date{\today}

\maketitle
Gapless electronic excitations observed in condensed matter systems are analogous to three types of fermionic-like elementary particles predicted by the standard model of particle physics, namely Dirac, Weyl, and Majorana fermions \cite{Palash, Armitage, Burkov_rev, Burkov_rev2, Jin_Hu, Weng_2016}. Compounds hosting such excitations include Dirac and Weyl semimetals with degenerate nodal points near their Fermi level $\varepsilon_F$. In addition to the two- and four-fold degenerate Weyl and Dirac semimetals, other classes of semimetals with three-, six-, and eight-fold degeneracy at high symmetry points within their Brillouin zone (BZ) have been predicted \cite{eight-fold-DDSM,Bernevig,ShouChengZhang-PRL,Burkov, ZrTe}.

For instance, owing to their chiral cubic crystallographic structure, exotic multifold topological excitations have been theoretically predicted \cite{ShouChengZhang-PRL, Burkov, RhSi-Hasan-PRL}, and recently observed in a family comprising transition metals combined with elements of the carbon group (or tetrels) as well as in aluminides including CoSi \cite{CoSi_Nature,CoSi_PRL} and $\beta$-RhSi \cite{CoSi-Hasan-Nat}, AlPt \cite{AlPt} and RhSn \cite{RhSn}.  These compounds belong to the non-centrosymmetric space group \textit{P}2$_1$3 (No. 198) for which the calculations predict Weyl nodes characterized by Chern numbers $C > 1$ at specific high symmetry points within their Brillouin zone (BZ) leading to several extended Fermi arcs on their surface.  Angle-resolved photoemission spectroscopy (ARPES) has verified the presence of chiral spin-1 fermions at the $\Gamma$-point and charge-2 fermions at the $R$-point of the BZ with $C=\pm2$ in CoSi and $\beta$-RhSi \cite{CoSi_Nature,CoSi_PRL,CoSi-Hasan-Nat}. Other types of fermions, the so called three-component fermions were predicted and experimentally observed in compounds crystallizing in the WC structure such as zirconium telluride (ZrTe)\cite{ZrTe} or molybdenum phosphide (MoP)\cite{MoP}. 
Owing to their non-trivial Chern numbers, chiral topological compounds are predicted to display novel effects like a magnetochiral anisotropy \cite{Morimoto} or the quantized circular photogalvanic effect \cite{deJuan, Flicker}.

Dirac semimetals, which, in contrast to Weyl systems preserve Kramers degeneracy, have also attracted a lot of attention recently. Compounds like Cd$_3$As$_2$ \cite{Cd3AS2_discovery} unveiled, for example, a possible novel mechanism of topological protection against backscattering \cite{ong} or, upon lifting time reversal symmetry and hence Kramers degeneracy, evidence for the chiral anomaly among field-induced Weyl nodes in Na$_3$Bi \cite{Na3Bi_discovery, Na3Bi_chiral}. This splitting of the Dirac into Weyl nodes has led to the prediction \cite{Kimchi} and subsequent observation \cite{Moll} of new cyclotron orbits involving Fermi arcs on the surface of Cd$_3$As$_2$ with an associated quantum Hall effect \cite{QHE_Cd3As2}. Topological nodal line (NL) systems, or systems where two bands cross forming closed lines within the BZ \cite{Burkov_NL,Fu_NL, NL_review,Chiu}, is another very active area of research. In general, spin-orbit coupling gaps the NLs unless they are protected by
some crystalline symmetry in addition to the inversion and time-reversal symmetry \cite{Burkov_NL,Fu_NL, NL_review}.
Prominent examples include the family of the ZrSiS compounds that display Dirac NLs (NLs resulting from the intersection of Dirac-like dispersing bands) at both symmorphic and nonsymmorphic positions \cite{Schoop} and the correlated Dirac NL iridate CaIrO$_3$ \cite{CaIrO3}.

Here, we report through group theory analysis and Density Functional Theory (DFT) calculations the existence of Dirac nodes and double Dirac nodes near high symmetry points, and very close to $\varepsilon_F$, in $\alpha$-RhSi, which adopts the orthorhombic MnP (B31) structure type (centrosymmetric space group \textit{Pnma}, No. 62). The intersection between the Dirac dispersions emerging from the double Dirac node leads to a symmetry protected Dirac nodal line. DFT calculations are confirmed via measurements of the dHvA-effect in flux grown crystals which unveils a Fermi surface whose topography is in good agreement with the predictions. For certain field orientations, the magnetic torque unveils an anomalous background suggesting transitions between distinct topological regimes, e.g. from Dirac to Weyl-like.
The topological character of the Rh silicides seems to survive the multiple structural phases observed in their binary phase-diagram.
\begin{figure}
	\centering
	\includegraphics[width = 8.6 cm]{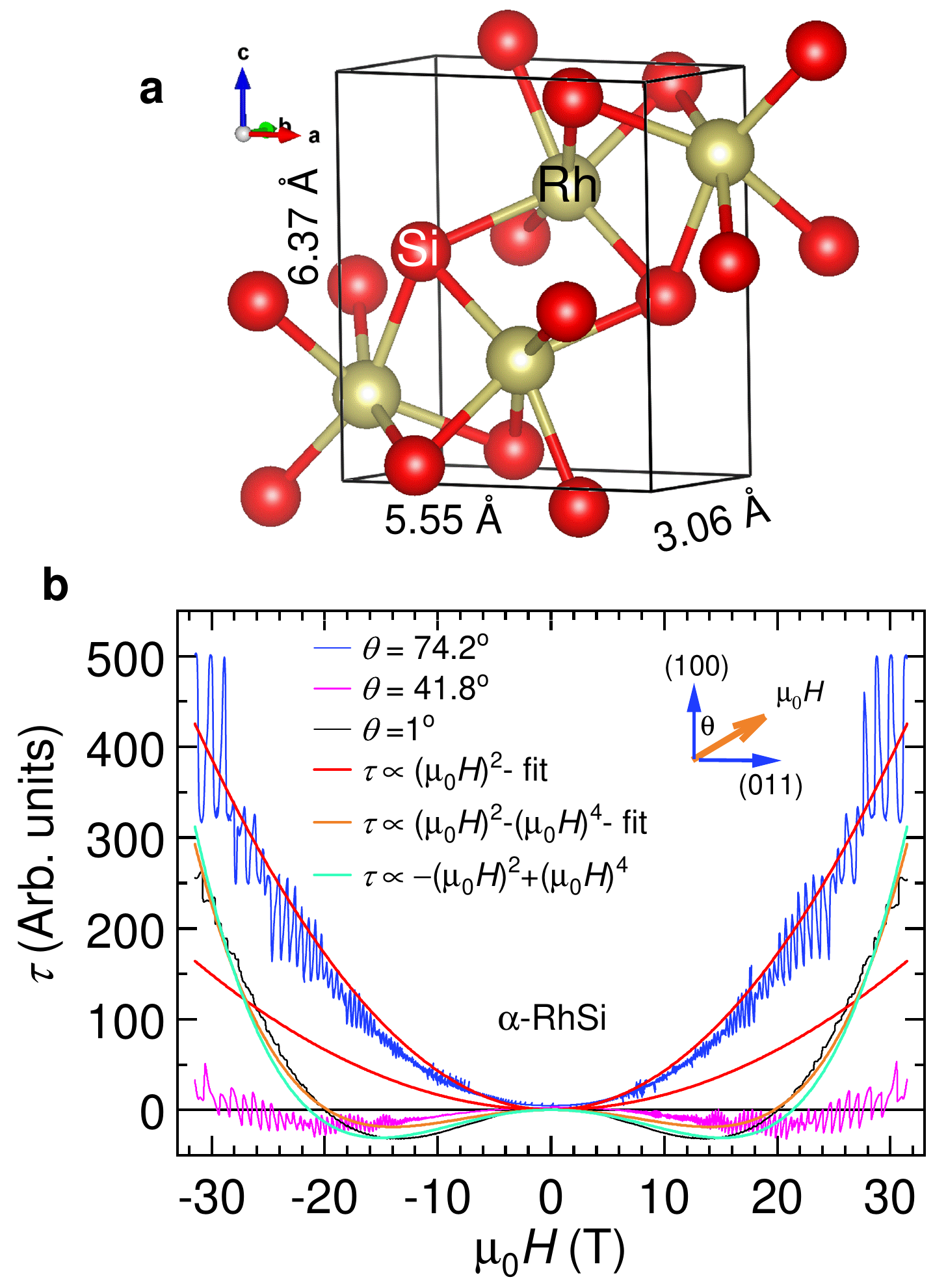}
	\caption{(Color online) (a)  Crystal structure of $\alpha$-RhSi. (b) Magnetic torque $\tau$ as a function of the applied magnetic field $\mu_0 H$ for several angles $\theta$ at $T = 350$ mK. $\theta$ is the angle between $\mathrm{\mu_0} H$ and the (100) plane of the crystal. Red, orange, and light green lines show different polynomial fittings.} \label{sketch}
\end{figure}

\begin{figure*}
	\centering
	\includegraphics[width=0.8\textwidth]{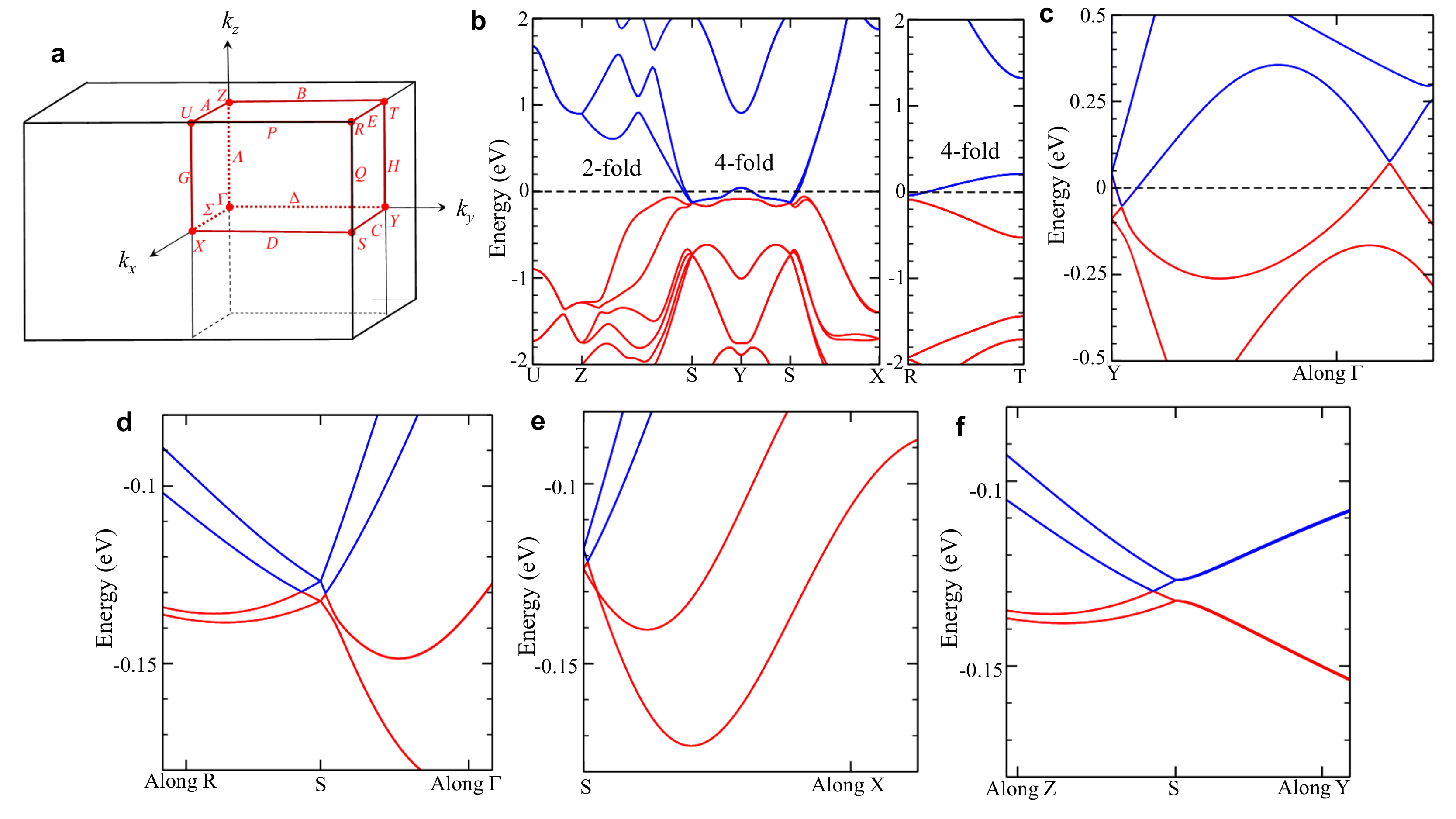}
	\caption{(Color online) (a) High-symmetry points on the orthorhombic BZ. (b) Electronic band structure along various high-symmetry directions where the 4-fold high-symmetry BZ boundary lines are indicated. (c),(d),(e), (f) Band structure along the various high-symmetry directions starting from the $S$ point, i.e. along the $Y\Gamma$, $SR$ and $S\Gamma$, $SX$, $ZS$ and $SY$ directions, respectively. According to our group-theory analysis regarding the connectivity of the $S$ point, there are band crossings along the $SR$ and $SX$ directions.
       }\label{bands}
\end{figure*}



Figure ~\ref{sketch}(a) displays the crystallographic structure of $\alpha$-RhSi along with its unit cell and lattice constants.
For details concerning sample synthesis, the DFT calculations \cite{PBE, QE-2009, ONCVPPHamann2013, Wannier90, Wannier902014, WIEN2K,SKEAF}, sample characterization, and measurements, see the Supplemental Information (SI) file \cite{SI}.
Fig.  S1 in the SI file shows the single crystal x-ray diffraction patterns along different \emph{hkl} planes, revealing
high crystallinity. Figure S2 displays an overall physical characterization of our $\alpha$-RhSi crystals as, for instance, the resistivity as a function of the temperature $T$ indicating a small residual resistivity $\rho_0 \sim$ 0.78 $\mu \Omega$ cm. In the same Fig. the magnetization reveals diamagnetic behavior over the entire temperature range 1.8 K $<T<$ 300 K, which is consistent with the absence of a Curie tail that one would associate with impurities.  From measurements of the Hall resistivity and of the magnetoresistivity at distinct $T$s we estimate the carrier concentrations $n_{e,h}$ and their mobilities $\mu_{e,h}$ through a two-band analysis. Mobilities and carrier concentrations are observed to increase as $T$ is lowered with this system behaving as a compensated semimetal. More importantly, at $T$ = 2 K one extracts high mobilities for holes, i.e. $\mu_h=3727$ cm$^2$/Vs and electrons $\mu_e=2491$ cm$^2$/Vs thus confirming the high quality of our single crystals. If instead, one evaluated the
mean transport mobility from the Hall-effect, as shown in Fig. S3 
one obtains  $\bar{\mu}_{tr} \simeq 6667$ cm$^2$/Vs at $T=2$ K, further confirming that our
flux grown crystals display a very low density of defects.

Figure \ref{sketch} (b) displays the magnetic torque $\tau = {(\mu_0H) \Delta M \sin2\theta}/2$, where $\Delta M$ is the anisotropy in the magnetization  $M_{c, ab} = \chi_{c,ab} (\mu_0H)$ ($\chi_{i}$ is the susceptibility along the $i$-axis) for a layered system, as a function of magnetic field $\mu_0 H$ and for three different angles $\theta$ between $\mu_0 H$ and the (100)-direction. $\tau$ is expected to behave as $\propto (\mu_0H)^2$ as illustrated by the fit (red line) to the $\tau(\mu_0H)$ trace collected at $\theta = 74.2^{\circ}$. The superimposed oscillatory component corresponds to the dHvA-effect resulting from the Landau quantization of the electronic spectrum. In contrast, at low angles the background of $\tau(\mu_0H)$ can only be fitted to non-physical combinations of quadratic and quartic terms. This sharp change in the behavior of $\tau(\mu_0H)$ cannot be reconciled with the diamagnetic response of this system. Therefore, it suggests the possibility of either topological phase-transitions or crossovers between distinct topological regimes induced by the Zeeman-effect. Below we provide the band structure calculations revealing a double Dirac structure whose nodes are very close in energy and hence susceptible to the energy scale of the external field.

In Fig.~\ref{bands} we  show the BZ, its high symmetry points, and the electronic band structure of orthorhombic $\alpha$-RhSi.
See also Fig. S4 \cite{SI} for the band structure depicted over the entire BZ.
Its various panels show that there are four-fold symmetries and the presence of several band crossings.
These crossings along various high symmetry directions can be understood from a symmetry analysis as presented next.
We will mainly focus on the $S$ high-symmetry point because there are band crossings near the Fermi level.
We also show that there should be a nodal line on the $k_y-k_z$ high-symmetry plane which is caused by these band crossings.
The close proximity in energy between these band-crossings, that are within the resolution of DFT,
requires an independent verification for their existence hence the subsequent group symmetry analysis.

The space group $Pnma$ (No. 62) is a non-symmorphic space group\cite{space-groups} with three screw-axis symmetry operations
of the general form using Seitz notation:
$\{2_{100}|{1\over 2}{1\over 2}{1\over 2}\}$,
$\{2_{010}|0{1\over 2}0\}$,
$\{2_{001}|{1\over 2}0{1\over 2}\}$,
and the three glide-plane symmetry operations: $\{m_{100}|{1\over 2}{1\over 2}{1\over 2}\}$,  $\{m_{010}|0{1\over 2}0\}$, $\{m_{001}|{1\over 2}0{1\over 2}\}$.
There are also time-reversal symmetry (TRS) and inversion symmetry which yield Kramers degeneracy.
The band-representations  of the double space group 62 at the various high symmetry points of the Brillouin zone for the Wyckoff position $8d:(x,y,z)$ (the case for $\alpha$-RhSi) are known \cite{Elcoro:ks5574,Bradlyn2017,PhysRevE.96.023310} and can be found in the Bilbao Crystallographic Server \cite{Bilbao}.

\begin{figure}[htb]
\begin{center}
\includegraphics[width = 8.6cm, trim=1cm 6.5cm 0.2cm 5.2cm, clip]{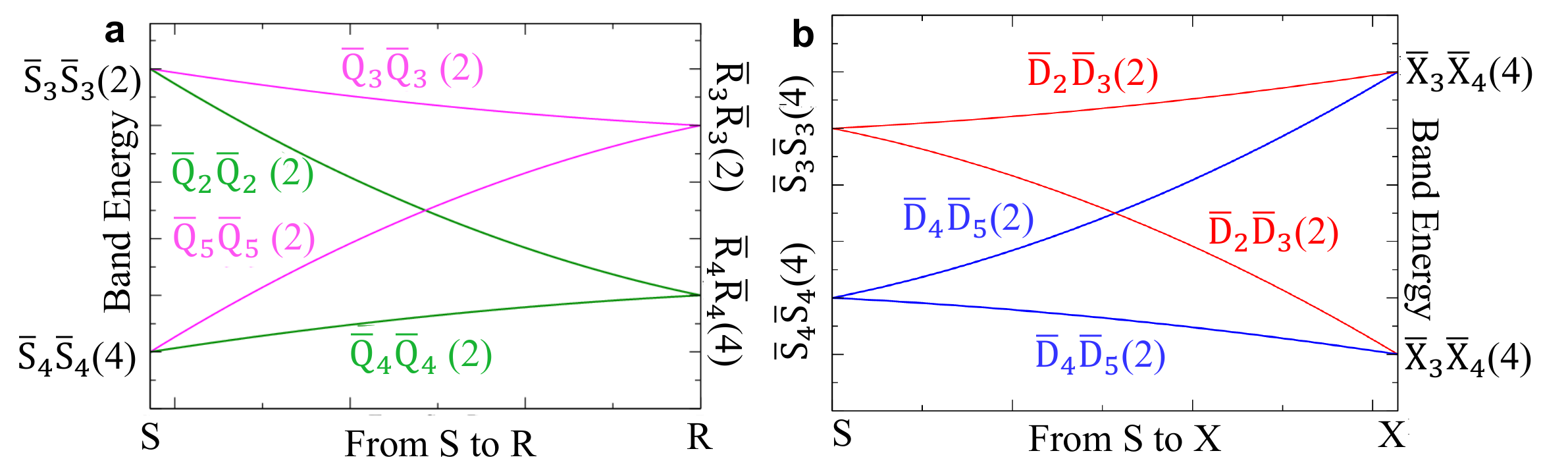}
\caption{(Color online) Band connectivity along the high-symmetry lines \emph{SR} (a) and \emph{SX} (b).} \label{crossings}
\end{center}
\end{figure}

The representations of  the subgroups of the high-symmetry points (HSPs) \emph{R}, \emph{S}, and \emph{U}, are direct sums of the two
four-dimensional (4D) representations that includes Kramers degeneracy.
Due to this additional degeneracy they consist of either a pair of the same irreducible representations (e.g., ${\bar R}_3{\bar R}_3$) or different irreducible representations (e.g., ${\bar R}_3{\bar R}_4$).
At the $\Gamma$ point there is no such additional degeneracy.
These are irreducible representations (IR) of the little group corresponding to each high-symmetry $k$-point and they are subduced along high-symmetry
lines that connects them. For example, along the \emph{RS} high-symmetry manifold, since the symmetry is lower than that of the high-symmetry points
that it connects, the IR ${\bar R}_3 {\bar R}_3(4)$ and ${\bar R}_4 {\bar R}_4(4)$ become reducible to
\begin{eqnarray}
  {\bar R}_3 {\bar R}_3(4) \to {\bar Q}_3 {\bar Q}_3(2) \oplus {\bar Q}_5 {\bar Q}_5(2), \\
  {\bar R}_4 {\bar R}_4(4) \to {\bar Q}_2 {\bar Q}_2(2) \oplus {\bar Q}_4 {\bar Q}_4(2).
\end{eqnarray}
Similarly, along the same high-symmetry line, the ${\bar S}_3 {\bar S}_3(4)$ and ${\bar S}_4 {\bar S}_4(4)$ representations are reducible to
\begin{eqnarray}
  {\bar S}_3 {\bar S}_3(4) \to {\bar Q}_2 {\bar Q}_2(2) \oplus {\bar Q}_3 {\bar Q}_3(2), \\
  {\bar S}_4 {\bar S}_4(4) \to {\bar Q}_4 {\bar Q}_4(2) \oplus {\bar Q}_5 {\bar Q}_5(2).
    \end{eqnarray}
Therefore, these IRs along these high symmetry lines connecting the two high symmetry points \emph{S} and \emph{R} have to be connected
as shown schematically in Fig.~\ref{crossings}. The compatible band connectivity, an example of which is shown in Fig.~\ref{crossings}(a), for any combination
of IRs at the two connected high-symmetry points, forces the intersection
of the bands belonging to different ${\bar Q}$-type IRs and this gives rise to a band crossing along these high-symmetry manifolds.
By carrying out the same analysis along the other high-symmetry lines that start
from $S$, we are concluding that these crossings form a
nodal line.
\begin{figure}[htb]
	\centering
	\includegraphics[width= 8.6 cm, trim=2.5cm 0cm 5cm 0cm, clip]{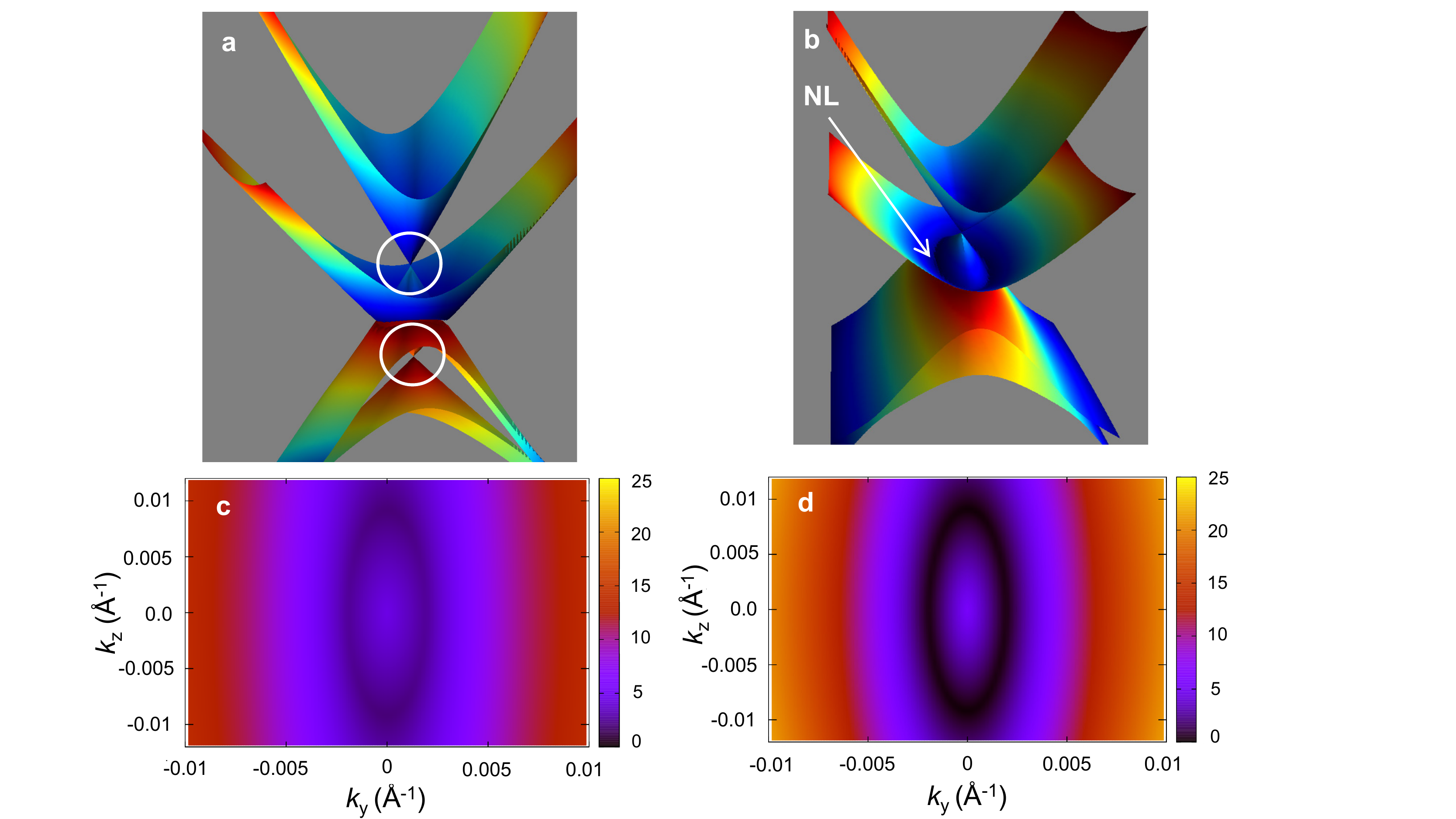}
	\caption{(Color online) (a) The band structure at the $S$ point as a function of the $(k_y,k_z)$. (b) Tilted band structure exposing the nodal line on the $k_y-k_z$ plane around the HSP \emph{S}. Circles and arrows indicate the Dirac nodes and the Dirac nodal line, respectively. (c) Constant energy contour for the lower band forming the nodal line. (d) Energy difference between the bands forming a nodal line. } \label{nodes} \label{nodes}
\end{figure}

A somewhat similar band crossing occurs when we consider the break-down of the IRs along the high symmetry lines which
start from the high-symmetry point \emph{S} and they connect the high-symmetry \emph{X} point (\emph{SX}-line or \emph{D} manifold) as seen in Fig.~\ref{crossings}(b).

The situation along the \emph{SY} direction is different. Along this symmetry line, the $ {\bar S}_3 {\bar S}_3(4)$ and $ {\bar S}_4 {\bar S}_4(4)$ at the \emph{S} point become:
\begin{eqnarray}
  {\bar S}_3 {\bar S}_3(4) \to {\bar C}_5 {\bar C}_5(4) \\
  {\bar S}_4 {\bar S}_4(4) \to  {\bar C}_5 {\bar C}_5(4).
    \end{eqnarray}
And the $ {\bar Y}_3 {\bar Y}_4(4)$ at the \emph{Y} point becomes
\begin{eqnarray}
  {\bar Y}_3 {\bar Y}_4(4) \to  {\bar C}_5 {\bar C}_5(4).
\end{eqnarray}
Therefore, the \emph{SY} line becomes a four-fold degenerate high-symmetry line.
\begin{figure}
\vspace{0.25 cm}
\begin{center}
	\includegraphics[width=8.6 cm]{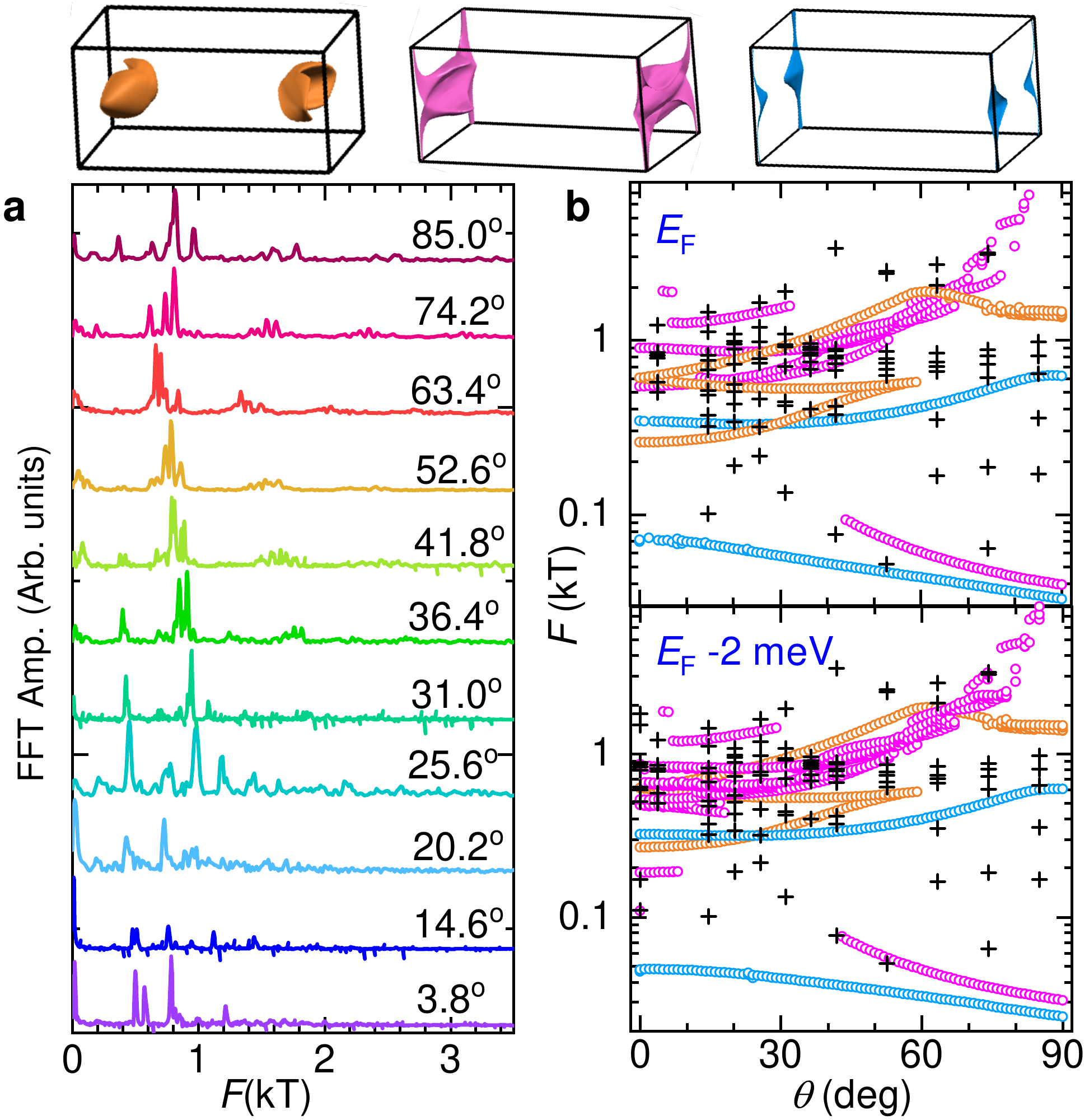}
	\caption{(Color online) (a) Fast Fourier transform spectra of the oscillatory component superimposed onto the magnetic torque $\tau$ for four different angles between the external field and the crystallographic (100)-direction. These traces were vertically displaced for clarity. The oscillatory data was collected at $T$=350 mK and is displayed in Fig. S3. (b) Angular dependence of the dHvA frequencies $F(\theta)$ for the position of the Fermi energy $\varepsilon_F$ yielded by the DFT calculations (upper panel) and with $\varepsilon_F$ displaced by just -2 meV (lower panel). Colored markers represent the dHvA frequencies obtained from the DFT calculations. Black crosses depict the position of the peaks observed in the FFT spectra. Notice the slightly better agreement between the experimental data and the frequencies resulting from the shift of $\varepsilon_F$. Top panels display the Fermi surface sheets according to DFT. Their color is chosen to match those of the markers depicting the associated dHvA frequencies.} \label{dHvA}
\end{center}
\end{figure}

Last, we consider the $\Delta$ manifold, i.e., the line connecting the $\Gamma$ and \emph{Y} high-symmetry points.
Along this direction the IRs at the Y point and $\Gamma$ point reduce as follows:
\begin{eqnarray}
  {\bar Y}_3 {\bar Y}_4(4) &\to& 2 {\bar \Delta}_5(2), \\
  {\bar \Gamma}_5(2) &\to&  {\bar \Delta}_5(2), \hskip 0.2 in  {\bar \Gamma}_6(2) \to  {\bar \Delta}_5(2).
\end{eqnarray}
Therefore, the ${\bar Y}_3 {\bar Y}_4(4)$ representation, which is irreducible at the \emph{Y} point, reduces into two
2D representations ${\bar \Delta}_5(2)$ which can directly connect to a 2D
${\bar \Gamma}_5(2)$ and a ${\bar \Gamma}_6(2)$ IRs at the $\Gamma$ point without or with
any number of band intersections. The latter seems to be the case in the band structure of $\alpha$-RhSi, namely, the two degenerate
bands at the \emph{Y} point separate along the \emph{Y} to $\Gamma$ direction and then they intersect each other twice along the
$\Delta$ direction before connecting to the $\Gamma$ point (See, Fig.~\ref{bands}(b)).
The \emph{SY} line is not the only four-fold high-symmetry manifold. The lines \emph{UZ} (\emph{A}-manifold), \emph{UX} (\emph{G}-manifold),
\emph{UR} (\emph{P}-manifold), and  \emph{RT} (\emph{E}-manifold) are all four-fold high-symmetry manifolds.

In the previous discussion  based on group representation at the high-symmetry points and from their connectivity through high-symmetry
lines, we demonstrated that there should be (symmetry-protected) band crossings along high-symmetry lines starting from the \emph{S} HSP.  This also became clear from Fig.~\ref{bands} where we noticed band crossings along $S\Gamma$, \emph{SX} and \emph{SR} lines. In fact, in Ref.~\onlinecite{Fu_NL}, it was
shown that a nonsymmorphic symmetry (screw axis)
protects a four-band crossing nodal line in systems having both inversion and time-reversal symmetries.
This applies directly to our case for
the $k_x=\pi$ plane (the plane where our nodal line lies)
which is invariant under the screw-axis operator $\{2_{100}|{1\over 2}{1\over 2}{1\over 2}\}$.
This implies that there is a symmetry protected nodal line around the \emph{S} point on the $k_y-k_z$ plane. This is illustrated in Fig.~\ref{nodes}(a) where on the $k_y-k_z$ plane there are two upside-down  Dirac cones intersecting and forming an elliptically shaped nodal line centered at the \emph{S} HSP, as shown in Fig.~\ref{nodes}(b). To clearly illustrate this further, Fig.~\ref{nodes}(c) shows a contour plot of the energy of the lower band, that is intersecting a higher energy band,
as a function of $k_y$ and $k_z$. In addition, Fig.~\ref{nodes}(d) gives a contour plot of the energy-difference between these two intersecting Dirac cones. Notice that the energy difference between the upper and lower Dirac cones at the $S$ point is only $\sim 5$ meV. As a result, because the Fermi velocity is $\hbar v_F \sim 1.13$ eV$\text{\AA }$ along the $k_y$ direction and 0.23 eV$\text{\AA }$ along the $k_z$-direction, the semimajor axis of the nodal line is about 0.01 $\text{\AA}^{-1}$ and the semiminor axis is about 0.002 $\text{\AA}^{-1}$. Therefore, a very dense $k$-point mesh was required to calculate this nodal line.

Finally, we confirm the validity of the DFT calculations through measurements of the angular dependence of the dHvA-effect. Figure ~\ref{dHvA} (a) displays the Fourier transform of the dHvA signal presented in Fig. S5(a), for several values of $\theta$. Each peak, observed at a given cyclotron frequency $F$, is related to a Fermi surface cross-sectional areas $A$ through the Onsager relation $F = \hbar A/2\pi e$. Therefore, in Fig. ~\ref{dHvA} (b), upper and lower panels,  we superimpose the $F$s extracted from the Fourier spectra in (a) (black crosses), with the $F$s expected from the Fermi surface cross-sectional areas (colored markers). Given the large number of orbits observed between 0.5 and 2.0 kT it is not possible to track their exact angular dependence, but one obtains a very good overlap between the predicted and the observed orbits in this frequency range.
Nevertheless, discrepancies are observed at lower frequencies that are likely  ascribable to the intrinsic errors associated with any specific DFT implementation.
In Fig. S5(b) we included the amplitude of the peaks observed in the FFT spectra as a function of the temperature, extracting effective masses ranging from 0.5 to 1.8 $m_e$ through the temperature damping factor in the Lifshitz-Kosevich formalism.

In summary, we have shown here, through a group theory analysis and DFT calculations, that the lower temperature phase of RhSi, i.e., the orthorhombic $\alpha$-RhSi, is also topological in character. It preserves roto-inversion symmetry and, therefore, instead of being a multifold Weyl fermion system as $\beta$-RhSi, it is a unique example of a Dirac semimetallic system, displaying single, as well as double Dirac nodes in close proximity to the Fermi level
(within 100 and 150 meV from $\varepsilon_F$). The double Dirac dispersions, emerging from the $S$-point with only 5 meV distance between the corresponding pair of Dirac points, intersect forming a nodal line in the $k_y-k_z$-plane protected by
the nonsymmorphic symmetry of $\alpha$-RhSi. This double Dirac structure and the associated Dirac nodal line is similar to what was proposed for the correlated iridate CaIrO$_3$ \cite{carter,CaIrO3}. However, in contrast to CaIrO$_3$, orthorhombic $\alpha$-RhSi does not display topologically trivial bands intersecting $\varepsilon_F$ \cite{CaIrO3}: all bands crossing $\varepsilon_F$ emerge directly from the Dirac nodes. Electronic correlations might be relevant also to $\alpha$-RhSi explaining perhaps the mild deviations observed between calculated and experimentally determined Fermi surface cross-sectional areas. Measurements of the de Haas-van Alphen effect in high quality flux grown single-crystals, reveal a Fermi surface whose topography is in good agreement with the DFT predictions, thus, validating the band structure calculations. The anomalous behavior of the magnetic torque observed for certain field orientations points to the possibility of transitions between distinct topological regimes as the Zeeman-effect displaces the spin-orbit split bands and lifts the Kramers degeneracy of the two Dirac dispersions.

L.B. is supported by DOE-BES through award DE-SC0002613. S.M. acknowledges support from the FSU Provost Postdoctoral Fellowship Program.
G.T.M. and J.Y.C. acknowledge support from  NSF-DMR-1700030. The National High Magnetic Field Laboratory acknowledges support from the
US-NSF Cooperative agreement Grant number DMR-1644779 and the state of Florida. The data in this manuscript can be assessed by requesting
it to the correspond authors.



%

\end{document}